\documentclass[11pt]{article}
\pdfoutput=1
\usepackage[latin1]{inputenc}
\usepackage{amsmath}
\usepackage{amsfonts}
\usepackage{amssymb}
\usepackage{pstricks}
\usepackage{amsthm}
\usepackage{mathrsfs}
\usepackage{amssymb}
\usepackage{cancel}
\usepackage{slashed}
\usepackage{graphicx}
\usepackage[font=small,labelfont=bf]{caption}
\usepackage{float}
\usepackage{csquotes}
\usepackage[bottom]{footmisc} 
\usepackage{tikz}
\usepackage{setspace}
\usepackage{bbding}
\usepackage{cite}
\usetikzlibrary{matrix,arrows,decorations.pathmorphing}
\usepackage{framed}
\usepackage{color}
\usepackage{wrapfig}
\definecolor{shadecolor}{RGB}{224,238,238}
\newcommand{\nn}{\nonumber}

\newcommand{\M}{\mathcal{M}}

\makeatletter

\@addtoreset{equation}{section}
\makeatother

\def\lsim{\;\raise0.3ex\hbox{$<$\kern-0.75em\raise-1.1ex\hbox{$\sim$}}\;}
\def\gsim{\;\raise0.3ex\hbox{$>$\kern-0.75em\raise-1.1ex\hbox{$\sim$}}\;}
\def\beq{\begin{equation}}   \def\eeq{\end{equation}}
\def\ba{\begin{array}}       \def\ea{\end{array}}
\def\bea{\begin{eqnarray}}   \def\eea{\end{eqnarray}}
\def\nn{\nonumber}
\def\nl{\newline}

\theoremstyle{definition} 
\date{\today}

\begin{document}

\begin{titlepage}
\begin{flushright}
LPT Orsay 16-07 \\
LUPM:16-004
\end{flushright}


\begin{center}

\begin{doublespace}

\vspace{1cm}
{\Large\bf A 750 GeV Diphoton Signal from a Very Light Pseudoscalar in the NMSSM} \\
\vspace{2cm}

{\bf{Ulrich Ellwanger$^{a,b}$ and Cyril
Hugonie$^c$}}\\
\vspace{1cm}
{\it  $^a$ Laboratoire de Physique Th\'eorique, UMR 8627, CNRS, Universit\'e de Paris-Sud,\\
Univ. Paris-Saclay, 91405 Orsay, France\\
\it $^b$ School of Physics and Astronomy, University of Southampton,\\
\it Highfield, Southampton SO17 1BJ, UK\\
\it $^c$ LUPM, UMR 5299, CNRS, Universit\'e de Montpellier, 34095
Montpellier, France }

\end{doublespace}

\end{center}
\vspace*{2cm}

\begin{abstract}
The excess of events in the diphoton final state near 750~GeV observed
by ATLAS and CMS can be explained within the NMSSM near the $R$-symmetry
limit. Both scalars beyond the Standard Model Higgs boson have masses
near 750~GeV, mix strongly, and share sizeable production cross sections
in association with b-quarks as well as branching fractions into
a pair of very light pseudoscalars. Pseudoscalars with a mass of
$\sim 210$~MeV decay into collimated diphotons, whereas pseudoscalars
with a mass of $\sim 500-550$~MeV can decay either into collimated diphotons
or into three $\pi^0$ resulting in collimated photon jets. Various such
scenarios are discussed; the dominant constraints on the latter scenario
originate from bounds on radiative $\Upsilon$ decays, but they allow for
a signal cross section up to 6.7~fb times the acceptance for collimated
multiphotons to pass as a single photon.

\end{abstract}

\end{titlepage}

\newpage
\section{Introduction}
\label{sec:intro}

In December 2015 the ATLAS and CMS collaborations have reported
excesses in the search for resonances decaying into pairs of photons for
diphoton invariant masses around 750~GeV~\cite{atlas750,cms750}.
In ATLAS, excesses appeared in the two $M_{\gamma\gamma}$ bins
710--750~GeV (14 events vs. 6.3 expected) and 750-790~GeV (9 events vs.
5.0 expected), with a local significance of $3.9\; \sigma$ (assuming a
large width of $\sim 45$~GeV; $3.6\; \sigma$ in the narrow width approximation). In 
CMS, excesses appear in the $M_{\gamma\gamma}$ bin 750--770~GeV
for photons in the EBEB category (5 events vs. 1.9 expected) and
EBEE category (6 events vs. 3.5 expected), but less in the bin 730-750~GeV
(4 events vs. 2.1 expected for photons in the EBEB category,
1 event vs. 4.0 expected for photons in the EBEE category, considered as
less sensitive). The local significance of the excesses is $2.6\; \sigma$
for CMS in the narrow width approximation.

The global significances of the signals of ${\cal O}(2-3\,\sigma)$ are
not overwhelming and  compatible with statistical fluctuations.
Still, the fact that the region of invariant diphoton masses is very
similar for ATLAS and CMS has stirred quite some excitement resulting
in a huge number of possible explanations. (The number of proposed
models exceeds the number of observed signal events.)

Fits to the combined data should, in principle, also consider the
informations from diphoton searches at 8~TeV~\cite{Aad:2015mna,Khachatryan:2015qba}
where a mild excess was observed by CMS. However, the extrapolation of
signal cross sections from 8 to 13 TeV depends on the assumed production
mechanism~\cite{Knapen:2015dap,Franceschini:2015kwy,Ellis:2015oso,
Bellazzini:2015nxw,Falkowski:2015swt}.
Assuming the production of a resonance around 750~GeV by gluon fusion (ggF),
combined fits to the signal cross sections at 13 TeV are in the range
2-10~fb~\cite{Knapen:2015dap,Franceschini:2015kwy,Ellis:2015oso,Falkowski:2015swt},
with slightly better fits and a larger signal cross section assuming a
larger width of 30-45~GeV~\cite{Knapen:2015dap,Franceschini:2015kwy,
Falkowski:2015swt}.

It is notoriously difficult to construct a consistent model for such a
resonance ``$X$'': Its production channel in proton proton collisions is typically
assumed to be ggF through loops of colored particles.
If these are the quarks of the Standard Model (SM), $X$ would decay into
them leaving little branching fraction for $X$-decays into $\gamma\gamma$,
which has to be generated by loop diagrams as well. 

Accordingly simple two Higgs doublet (or MSSM) extensions of the Standard Model,
which could contain a resonance $X$ near
750~GeV~\cite{Angelescu:2015uiz,Becirevic:2015fmu,Han:2015qqj,Huang:2015rkj,
Moretti:2015pbj,Badziak:2015zez,Hall:2015xds,Bizot:2015qqo,
Ma:2015xmf,Han:2016bus,Hernandez:2016rbi,
Han:2016bvl,King:2016wep,Bertuzzo:2016fmv},
require additional scalars or vector-like fermions whose loops generate the
coupling of $X$ to
gluons and/or $\gamma\gamma$ (unless $R$-parity is broken~\cite{Ding:2015rxx,Allanach:2015ixl}). 
Large Yukawa couplings are required for a sufficiently
large cross section, which risk to generate new hierarchy problems/Landau singularities
(unless compositeness is invoked). Also in the
Next-to-Minimal supersymmetric extension of the Standard Model (NMSSM) it has been
argued~\cite{Tang:2015eko,Wang:2015omi,Chao:2016aer} that additional vector-like quark
superfields have to be introduced. In~\cite{Badziak:2016cfd} a two-step decay cascade involving
the two pseudoscalars of the NMSSM with masses of about 750~GeV and 850~GeV has been
proposed which requires, however, to tune the corresponding mixing angle close to~0.

A different approach towards an explanation of the diphoton events is to
consider that a single photon in the detector can represent a collimated
bunch of photons (typically two of them) which originate from a single very light state,
for instance a light pseudoscalar $A$~\cite{Knapen:2015dap,Agrawal:2015dbf,Chang:2015sdy,
Chala:2015cev,Bi:2015lcf,Aparicio:2016iwr,Dasgupta:2016wxw,Chiang:2016eav,Arcadi:2016dbl}.
Then the observed processes correspond to an initial resonance
$X$ decaying into a pair  $A\,A$, where $M_A$ must be well below 1~GeV
for the resulting photons to be sufficiently collimated (see below).
This scenario opens the possibility to explain the diphoton events in different
models which can accomodate resonances $X$ and a light pseudoscalar $A$. In
this paper we show that the simple $\mathbb{Z}_3$-invariant NMSSM belongs
to this class of models. (This has also been observed in~\cite{Domingo:2016unq}.)

In the NMSSM (see~\cite{Maniatis:2009re,Ellwanger:2009dp} for reviews),
two CP-even Higgs states beyond
the Standard Model-like Higgs (subsequently denoted as
$H_{SM}$) can play the role of a resonance $X$. In terms of weak eigenstates,
a singlet-like state $S$ can have a large coupling to a pair of mostly
singlet-like pseudo\-scalars $A_1$, originating from a cubic singlet self
coupling $\kappa$ in the superpotential (see below). However, a coupling to
quarks or gluons inside protons has to be induced by a mixing of $S$ with one of
the two SU(2) doublet-like Higgs states. If this state is $H_{SM}$,
the mixing reduces the couplings of $H_{SM}$ to SM particles (notably
$W^\pm$ and $Z$) and is severely constrained~\cite{Falkowski:2015swt}
by the measured signal rates by
ATLAS and CMS~\cite{ATLAS-CMS:2015kwa}. An alternative is that $S$ mixes strongly with
the other ``MSSM''-like CP-even state $H$. Then the physical eigenstates
-- preferably both of them with masses near 750~GeV -- can profit from
an enhancement of the couplings of $H$ to $b$-quarks by $\tan\beta$,
leading to sufficiently large signal cross sections into the $A_1A_1$ (and
hence diphoton) final state via associated production with $b$-quarks.
Given the diphoton mass resolution of the detectors and the slightly
preferred large width of the excess it is clear that two (narrow) CP-even states
near 750~GeV, mixtures of $H$ and $S$, can also provide a good fit to the data.
(A similar scenario has been discussed in~\cite{Cao:2016cok}.)
For one of the benchmark points presented below (BP1) the signal originates,
however, from one CP-even state only, the other one being significantly
heavier.

A light pseudoscalar can appear in the NMSSM in the form of a
pseudo-Goldstone boson (PGB). A priori two global symmetries can lead to
such PGBs: First, a Peccei-Quinn symmetry emerges in the limit $\kappa \to
0$~\cite{Ellwanger:2009dp,Miller:2003ay,Hall:2004qd,Barbieri:2007tu}.
However, $\kappa \neq 0$ is required for the couplings of the heavy Higgs states
to $A_1A_1$. Second, the scalar potential
of the NMSSM is invariant under an $R$-symmetry~\cite{Ellwanger:2009dp,Dobrescu:2000yn,
Dermisek:2006wr,Morrissey:2008gm} if the soft
supersymmetry breaking trilinear couplings $A_\lambda$ and $A_\kappa$
vanish, leading to a PGB due to its spontaneous breakdown by the
phenomenologically required vacuum expectation values. We find indeed,
that the interesting part of the parameter space of the NMSSM corresponds
to small values of $A_\lambda$ and $A_\kappa$. However, since the $R$-symmetry
is broken by radiative corrections to the scalar potential involving the necessarily
non-vanishing gaugino masses and trilinear couplings $A_t$ and $A_b$, it
helps only partially to explain a very light pseudoscalar $A_1$. Still, it
represents a ``go-theorem'' showing
that a standard supersymmetric extension of the SM -- without additional
vector-like quarks and/or leptons -- could explain the observed diphoton
excess.

Different assumptions on the mass of $A_1$ can be made.
For one set of scenarios we assume $M_{A_1} \sim 210$~MeV, just below the
$2\,\mu$ threshold. These scenarios lead to visibly displaced vertices
from the $A_1\to \gamma\gamma$ decays. For a large value of the NMSSM
trilinear coupling $\kappa \sim 1.65$, the signal can originate from a
single Higgs state near 750~GeV. For smaller values of $\kappa$, the signal
can originate from two Higgs states with masses near 750~GeV.
For another set of scenarios we assume $M_{A_1} \sim 510-550$~MeV, not far from the
$\eta$ mass. For $M_{A_1}$ near 550~MeV, $A_1$ mixes with the $\eta$ meson and
inherits its decays into $\gamma\gamma$ and $3\pi^0$; the latter lead to
photon-jets. The average separation in rapidity of the diphotons and the two leading
photons from $3\pi^0$ will be studied.
For $M_{A_1}$ near 510~MeV, constraints from searches for
radiative $\Upsilon(1S)$ decays into $\gamma + \eta$ by CLEO~\cite{Athar:2007hz}
are alleviated, but estimates of the $A_1$ decay widths are more uncertain.
But in both cases the $A_1$ life time is short enough avoiding
macroscopically displaced vertices,
and two Higgs states near 750~GeV can generate a signal.

In the next section we describe with the help of analytic approximations
to the mass matrices (including only the dominant radiative corrections)
which region in the parameter space of the NMSSM can generate the
diphoton events. In section~3 we discuss various
constraints from low energy physics on light pseudoscalars, and discuss
separately the different scenarios. Benchmark points are presented
with the help of the public Fortran code 
NMSSMTools~\cite{Ellwanger:2004xm,Ellwanger:2005dv}.
For the different $A_1$ masses we study the average separation in rapidity
of the diphotons and the two leading photons from $3\pi^0$, which allows
to estimate the corresponding acceptances.
In the final section~4 we summarize and
discuss possible alternative signatures, which could help to distinguish
different scenarios
if the excess survives the next runs of the LHC.

\section{Parameter regions with diphoton-like events at 750~GeV in the NMSSM}

We consider the CP-conserving $\mathbb{Z}_3$-invariant NMSSM.
The superpotential of the Higgs sector reads in terms of hatted superfields
\beq
W_{\text{Higgs}}=\lambda\hat{S}\hat{H}_u\cdot \hat{H}_d
+\frac{\kappa^3}{3}\hat{S}^3\; .
\label{2.1e}
\eeq
Once the real component of the singlet superfield $\hat S$ develops a
vacuum expectation value (vev) $s$, the first term in $W_{\text{Higgs}}$
generates an effective $\mu$ term
\beq
\mu= \lambda\, s\; .
\eeq
The soft SUSY--breaking terms consist of mass terms for the
gaugino, Higgs and sfermion fields
 \bea
-{\cal L}_\mathrm{\frac12}\!&\!=\!&\! \frac{1}{2} \bigg[ 
 M_1 \tilde{B}  \tilde{B}
\!+\!M_2 \sum_{a=1}^3 \tilde{W}^a \tilde{W}_a 
\!+\!M_3 \sum_{a=1}^8 \tilde{G}^a \tilde{G}_a   
\bigg]+ \mathrm{h.c.}\; , \nn \\ 
 -{\cal L}_\mathrm{0} \!&\!=\!&\! 
m_{H_u}^2 | H_u |^2 + m_{H_d}^2 | H_d |^2 + 
m_{S}^2 | S |^2 +m_Q^2|Q^2| + m_{T}^2|T_R^2| \nn \\ &
&+\,m_B^2|B_R^2| +m_L^2|L^2|+m_\mathrm{\tau}^2|\tau_R^2|\; ,
\eea
as well as trilinear interactions between the sfermion and the Higgs
fields, including the singlet field 
\bea
-{\cal L}_\mathrm{tril} \!&\!=\!&\! 
 \Bigl( h_t A_t\, Q\cdot H_u\, T_R^c
+ h_b  A_b\, H_d \cdot Q\, B_R^c + h_\tau A_\tau \,H_d\cdot L
\,\tau_R^c 
\nn \\ \!& &\!
+\,  \lambda A_\lambda\, H_u \cdot H_d \,S +  \frac{1}{3} \kappa 
 A_\kappa\,  S^3 \Bigl)+ \mathrm{h.c.}\;.
\eea

The tree level scalar potential can be found in~\cite{Ellwanger:2009dp}, from which the
$3\times 3$ mass matrices in the CP-even and CP-odd sectors can be
obtained.
Once the soft Higgs masses are expressed in terms of $M_Z$, $\tan\beta$ and
$s$ using the minimization equations of the potential, the mass matrices
depend on the six parameters
\beq
\lambda, \quad\kappa, \quad\tan\beta=\frac{v_u}{v_d},\quad \mu, \quad A_\lambda\quad
\text{and}\quad A_\kappa\; .
\label{2.5e}
\eeq

Initially, the CP-even mass matrix $\M^2_S$ is obtained in the basis of the real
components\break
 $\left(H_{d,r}, H_{u,r}, S_r\right)$ of the complex scalars $\left(H_{d}, H_{u}, S\right)$
after expanding around the vevs $v_d,\ v_u$ and $s$.
It is convenient, however, to rotate $\M^2_S$ by an angle $\beta$ in the doublet sector sector
into $\M'^2_S$ in the basis $H_{SM}',H',S_r$:
\beq
\M'^2_{S} = R(\beta) \M^2_S R^{\cal T}(\beta)\; , \qquad
R(\beta) = \left(\ba{ccc} \cos\beta & \sin\beta & 0 \\ 
\sin\beta & -\cos\beta & 0\\
0 & 0 & 1 \ea\right)\; .
\eeq
The advantage of this basis is
 that only the component $H_{SM}'$ of the Higgs doublets acquires
a vev $v$ and that, for typical parameter choices, it is nearly diagonal:
$H_{SM}'$ has SM-like
couplings to fermions and electroweak gauge bosons,
the  heavy doublet field $H'$ is the CP-even partner of the MSSM-like
CP-odd state $A_{MSSM}$,
while $S_r$ remains a pure singlet. The mass matrix
$\M'^2_S$ in the basis $(H_{SM}',H',S_r)$ has the elements
\begin{eqnarray}
\M'^2_{S,11}&=& M^2_Z\cos^2 2\beta + \lambda^2 v^2 \sin^2 2\beta + \sin^2\beta \Delta_{\text{rad}}\; ,
\nn \\
\M'^2_{S,12}&=& \sin 2\beta\left( \cos 2\beta \left(M_Z^2-\lambda^2 v^2\right) 
- \frac{1}{2}\Delta_{\text{rad}}\right)\; ,\nn \\
\M'^2_{S,13}&=& \lambda v \left(2\mu -(A_{\lambda} + 2\kappa s) \sin 2\beta\right)\; ,\nn\\
\M'^2_{S,22}&=& M_A^2+\left(M_Z^2-\lambda^2 v^2 \right)\sin^2 2\beta + \cos^2\beta \Delta_{\text{rad}}
\; ,\nn\\
\M'^2_{S,23}&=& \lambda v(A_{\lambda} + 2\kappa s) \cos 2\beta\; ,\nn \\
\M'^2_{S,33}&=& \lambda A_\lambda\frac{v^2}{2s} \sin 2\beta 
+ \kappa s\left(A_\kappa + 4\kappa s\right)\; ,
\label{2.7e}
\end{eqnarray}
where $v^2=2M_Z^2/(g_1^2+g_2^2)\sim (174\ \mathrm{GeV})^2$ and
\beq
M_A^2=\frac{2\mu}{\sin 2\beta}(A_\lambda+\kappa s)
\eeq
is the mass squared of the MSSM-like CP-odd state $A_{MSSM}$.
$\Delta_{\text{rad}}$ denotes the dominant radiative corrections due to top/stop loops,
\beq
 \Delta_{\text{rad}}=\frac{3 m_t^4}{4\pi^2 v^2}
\left(\ln\left(\frac{m_{ST}^2}{m_t^2}\right)+\frac{X_t^2}{m_{ST}^2}
 \left(1-\frac{X_t^2}{12 m_{ST}^2}\right)\right)
\label{2.9e}
\eeq
where $m_{ST}^2=m_Q m_T$  and  $X_t = A_{t} -\mu/\tan\beta$.

As discussed in the introduction, we intend to describe the diphoton signal at $\sim 750$~GeV
by a mixture of the two states $H'$ and $S_r$. Then, both diagonal matrix elements
$\M'^2_{S,22}$ and $\M'^2_{S,33}$ should have values close to $(750~\text{GeV})^2$.
Furthermore we will be interested in the $R$-symmetry limit
$A_\lambda,\ A_\kappa\to 0$. This implies the relations (for $\tan^2\beta \gg 1$)
\beq
\M'^2_{S,22}\sim M_A^2 \sim \frac{2\mu\kappa s}{\sin 2\beta} \sim
\frac{\kappa}{\lambda} \mu^2\tan\beta \sim (750~\text{GeV})^2
\label{2.10e}
\eeq
and
\beq
\M'^2_{S,33} \sim (2\kappa s)^2 \equiv 4\left(\frac{\kappa}{\lambda}\right)^2
\mu^2 \sim (750~\text{GeV})^2\; .
\label{2.11e}
\eeq

The matrix element inducing $H'-S_r$ mixing is given by
\beq
\M'^2_{S,23} \sim 2\kappa v\mu\; ,
\label{2.12e}
\eeq
and the matrix element inducing $H'_{SM}-S_r$ mixing by
\beq
\M'^2_{S,13} \sim  2\lambda v\mu \; .
\label{2.13e}
\eeq

Next we turn to the CP-odd sector. The $3\times 3$ CP-odd mass matrix contains
always a Goldstone boson which will be eaten by the $Z$ boson. The remaining
CP-odd states are a singlet $A_S$, and the ``MSSM''-like SU(2)-doublet $A_{MSSM}$.
In the basis $(A_{MSSM}, A_S)$, in the $R$-symmetry limit $A_\lambda,\ A_\kappa\to 0$,
 the CP-odd mass matrix is given by
\beq
\M_A^2 = \frac{2\kappa\mu}{\sin 2\beta}
 \left(\ba{cc} s & -v \sin 2\beta \\ 
-v \sin 2\beta & \frac{v^2}{s} \sin^2 2\beta  \ea\right)\; .
\label{2.14e}
\eeq
Obviously $\M_A^2$ has a vanishing eigenvalue $M_{A_1}=0$, and is diagonalised by
an angle $\alpha$ with (for $\tan^2\beta \gg~1$)
\beq
\sin\alpha  \approx \frac{2 v}{s\tan\beta}\; .
\label{2.15e}
\eeq

An important quantity will be the (reduced) coupling $X_d$ of $A_1$ to down quarks
and leptons, which is obtained
through the mixing of $A_S$ with $A_{MSSM}$.
Since the reduced coupling of the MSSM-like state $A_{MSSM}$
is given by $\tan\beta$, one obtains
\beq
X_d \sim \sin\alpha\tan\beta \sim \frac{2v}{s}\equiv \frac{2\lambda v}{\mu}\; .
\label{2.16e}
\eeq

Radiative corrections to the tree level potential and hence to
the CP-odd mass matrix include terms
proportional to the electroweak gaugino masses $M_1$ and $M_2$, and
terms proportional to
the soft SUSY breaking trilinear couplings $A_t$ and $A_b$.
These corrections break the $R$-symmetry present for $A_\lambda,\ A_\kappa\to 0$,
which is expected since $A_\lambda,\ A_\kappa = 0$ is not invariant under scale
transformations. Hence, depending on the scale where $A_\lambda,\ A_\kappa = 0$
is assumed to hold, $A_1$ is a pseudo-Goldstone boson with a mass of
typically a few GeV. For $A_\kappa$ small, but $\neq 0$ one can obtain $M_{A_1}\sim 210$~MeV
or $M_{A_1}\sim 510-550$~MeV as it will be assumed in the next section.

Finally we note that, for the parameter region considered below,
 the dominant contribution to the coupling of $A_1$ to scalars
originates from the quartic coupling $\sim \kappa^2 |S|^4 \to
2\kappa^2 S_r^2 A_1^2$. After shifting $S_r$ by its vev $s$ one obtains
\beq
g_{S A_1 A_1} \sim \sqrt{2} \kappa^2 s\; .
\label{2.17e}
\eeq

Next we observe that eqs. \eqref{2.11e} and \eqref{2.16e}
allow to express $\kappa$ in terms of $X_d$: From \eqref{2.11e} one
finds
\beq
750~\text{GeV}\sim 2\kappa s = 2\kappa\frac{\mu}{\lambda}
=\frac{4\kappa v}{X_d}
\eeq
where \eqref{2.16e} was used in the last step. Inserting $v\sim 174$~GeV
one obtains
\beq
\kappa \sim 1.1\, X_d\; .
\label{2.19e}
\eeq
In the next section, for the scenarios with $M_{A_1} \sim 510-550$~MeV,
we will obtain upper bounds on $X_d$ from upper bounds for the
$BR(\Upsilon(1S)\to\gamma \eta)$ from CLEO~\cite{Athar:2007hz}. These
will thus imply upper bounds on $\kappa$ according to \eqref{2.19e}.
On the other hand a large signal rate, generated by a mixture of the
states $H'$ and $S$ decaying into $A_1 A_1$, requires $g_{S A_1 A_1}$
to be as large as possible. Accordingly $X_d$ and $\kappa$ should saturate
corresponding upper bounds.

If the 750~GeV signal is generated by a superposition of signals of
two nearby physical states formed by the $H'-S_r$ system, their
mass splitting should not be too
large, preferably of ${\cal O}(20\ \text{GeV})$. Then the
matrix element $\M'^2_{S,23}$ given in \eqref{2.12e} should be as small
as possible. With $\kappa$ already determined, this implies $\mu$ as
small as possible, preferably close to the lower bound $\sim 100$~GeV
from the LEP lower bound on higgsino-like charginos. Then
\eqref{2.16e} requires that $\lambda$ is relatively small.
(Simultaneously, this avoids a strong push-down effect on the mass
of the SM-like Higgs boson from $H_{SM}-S_r$ mixing, which is induced by
the matrix element $\M'^2_{S,13}$ given in \eqref{2.13e}.)
Finally the condition \eqref{2.10e} on $M_A^2$ fixes $\tan\beta \approx 15$.

The remaining NMSSM parameters in \eqref{2.5e} are $A_\lambda$ and $A_\kappa$.
Both $R$-symmetry breaking para\-meters have an impact on the mass of the
pseudo-Goldstone boson $A_1$. We find that one can chose small
values of $A_\lambda$ and
$A_\kappa$ such that $M_{A_1}$ assumes the desired value;
due to radiative corrections to the scalar potential the precise value
of $A_\kappa$ depends on the other $R$-symmetry breaking parameters
$M_1$, $M_2$, $A_t$ and $A_b$. 
Herewith all NMSSM parameters are nearly uniquely determined.

\section{Viable scenarios with a light NMSSM pseudoscalar}

As discussed in the introduction we will study scenarios with different
values of the mass of a light pseudoscalar, denoted subsequently by 
$M_{A_1}$. Constraints on such a light NMSSM pseudoscalar with a mass below
$\sim 1$~GeV have been discussed previously
in~\cite{Dobrescu:2000jt,Hiller:2004ii,Domingo:2007dx,Domingo:2008bb,
Domingo:2008rr,Andreas:2010ms,Jaeckel:2015jla,Domingo:2015wyn}. Strong constraints
originate from the mediation of FCNCs. Assuming minimal flavour
violation, flavour violating couplings of $A_1$ still originate from SUSY
loops involving stops, sbottoms and charginos and
depend on the corresponding masses and trilinear couplings like $A_t$.
These contribute notably to $B$-physics observables like
$B_s \to \mu^+\mu^-$, 
$\Delta M_d$ and $\Delta M_s$. We have implemented the computation of
these and many more $B$-physics observables and some $K$-physics observables
in the code NMSSMTools~\cite{Ellwanger:2004xm,Ellwanger:2005dv} following the
update in~\cite{Domingo:2015wyn} and checked that, for the scenario presented here,
the constraints are satisfied due to the mostly singlet-like nature of $A_1$
and the relatively heavy SUSY spectrum. 

For $M_{A_1}$ near 210~MeV, additional strong constraints originate
from rare flavour changing processes $K^\pm \to \pi^\pm e^+ e^-$.
(In~\cite{Andreas:2010ms}
it has been argued that the corresponding constraints exclude scenarios with
$M_{A_1} \lsim 210$~MeV, where the branching fraction of $A_1$ into $e^+e^-$ is sizeable.)
We have verified the assertion in~\cite{Domingo:2016unq} that, for suitable
choices of soft SUSY breaking parameters, the coupling $C_A$ responsible
for these processes (see~\cite{Hiller:2004ii})
can be arbitrarily small\footnote{We thank F. Domingo for help for
this calculation.}. 
Light pseudoscalars have been searched for in radiative $\Upsilon(1S)$ decays
by CLEO in~\cite{Love:2008aa}; these are also verified by NMSSMTools\_4.9.0
and satisfied by the benchmark points given below.

Due to the mostly singlet-like
nature of $A_1$, its contributions to the muon anomalous magnetic moment
are negligibly small. However, for $\tan\beta \sim 15$ and assuming
relatively light slepton masses of 300~GeV, the scenarios below can reduce
the discrepancy between the measured value and the Standard Model to
an acceptable $2\,\sigma$ level.

Further constraints stem from possible $A_1$ production in $Z$ and $H_{SM}$
decays. The relevance of bounds on light pseudoscalars (or axion-like
particles) from searches for $Z\to \gamma\gamma$ at LEP
(where a photon can correspond to a bunch of collimated photons)
has been investigated in~\cite{Jaeckel:2015jla}. These bounds constrain
the loop-induced coupling $g_{ZA\gamma}$. This coupling is also constrained
by the upper bound on $BR(Z\to \eta \gamma) < 5.1\times 10^{-5}$~\cite{Agashe:2014kda}.
We have checked that in our cases this coupling is about four
orders of magnitude below the bounds derived from~\cite{Jaeckel:2015jla,Agashe:2014kda}.
Searches for $H_{SM} \to A_1A_1 \to 4\gamma$ have been undertaken by
ATLAS using 4.9~fb$^{-1}$ of integrated luminosity at 7~TeV c.m.~energy
in~\cite{ATLAS-CONF-2012-079} for $M_{A_1} < 400$~MeV. 
One can assume that the corresponding
upper bound on $BR(H_{SM}\to A_1A_1)\times BR(A_1\to\gamma\gamma)^2 \lsim
6.6\times 10^{-3}$ applies to our scenario as well, which leads to
$BR(H_{SM}\to A_1A_1) \lsim 1.7\times 10^{-2}$. If $A_1$ imitates a single
photon, bounds on $BR(H_{SM}\to \gamma\gamma)$ should be respected.
In our scenarios we require
$BR(H_{SM}\to A_1A_1) \lsim 5\times 10^{-4}$, hence these constraints are
well satisfied. Notably this small branching fraction has no impact
on the measured signal rates of $H_{SM}$ into the other Standard Model channels,
which agree well with the Standard Model predictions.

Additional constraints depending on $M_{A_1}$ will be
discussed in the corresponding subsections below.

\subsection{$M_{A_1}$ near 210~MeV}

For a light $A_1$, too light for hadronic final states ($M_{A_1}$ below
$3M_{\pi}$), the possible decays are into
$\mu^+\mu^-$, $e^+e^-$ and the loop induced decay into $\gamma\gamma$.
The couplings of $A_1$ to Standard Model fermions are obtained via
mixing with $A_{MSSM}$ as discussed in eqs.~\eqref{2.14e} and
\eqref{2.15e} in the previous section, and lead to a reduced coupling
of $A_1$ to leptons $\sim X_d \sim \kappa$, see \eqref{2.19e}.
These couplings determine also
the partial width into $\gamma\gamma$.
For a sizeable branching fraction into $\gamma\gamma$, the decay into
$\mu^+\mu^-$ must be kinematically forbidden. On the other hand,
for $M_{A_1} < 200$~MeV the remaining decays into $e^+e^-$ and
$\gamma\gamma$ lead generically to a too small total width
implying, for a boosted $A_1$ with an energy of about 375~GeV, a decay
length larger than the size of the detectors (unless $A_1$ mixes strongly with
$\pi^0$ as discussed in~\cite{Domingo:2016unq}). However, for $M_{A_1}$
very close to $2 m_\mu$, the loop contribution of muons to the width
$\Gamma(A_1 \to \gamma\gamma)$ reaches a maximum. It is given by (neglecting
all other contributions; see, e.g.,~\cite{Ellis:1975ap})
\beq
\Gamma(A_1\to \gamma\gamma)|_\text{muons}
= \frac{G_\mu \alpha_{em}^2 M_{A_1}^3}{128\sqrt{2}\pi^3} X_d^2
\left|A_{1/2}^A(\tau)\right|^2
\eeq
with $\tau = M_{A_1}^2/(4 m_\mu^2)$ and, for $\tau \leq 1$,
\beq
A_{1/2}^A(\tau_f)=2\tau^{-1} \arcsin^2\sqrt{\tau}\; ;
\eeq
accordingly it increases with $M_{A_1} \to 2 m_\mu$ (remaining finite
for $M_{A_1} = 2 m_\mu$). We find that, for $M_{A_1}$ near or
slightly above 210~MeV,
the partial width $\Gamma(A_1\to \gamma\gamma)$ dominated by the muon
contribution is large enough to dominate the width $\Gamma(A_1\to e^+e^-)$
leading to a $BR(A_1\to \gamma\gamma) \sim 74\%$.

The total $A_1$ width depends then essentially on its reduced coupling to muons
$X_d$ related to $\kappa$
via~\eqref{2.19e}. First we consider a scenario with a total
width of $\sim 1.7\times 10^{-13}$~GeV, leading to a decay length of $A_1$ for an energy
of 375~GeV of about 2~m. Given that the distance of the EM
calorimeter cells to the interaction point is larger than 1.3~m for the ATLAS
and CMS detectors (depending on the angle $\eta$), one can estimate that
somewhat more than 60\% of all pseudoscalars decay before the EM
calorimeter cells.

This scenario requires $\kappa \gsim 1.65$, in which case
$\kappa$ runs into a Landau singularity at about 400~TeV where
the NMSSM would require a UV completion (e.g. GMSB). Then
a single Higgs state near 750~GeV is
able to generate a visible signal. (The second Higgs state is heavier near 1~TeV and
has a significantly smaller production cross section. A scenario where
a single Higgs state near 750~GeV is responsible for the signal and another
Higgs state is far below 750~GeV is not possible: Then the lighter state
would generate a larger signal, which is excluded.) For a large enough
production cross section of the state near 750~GeV from its coupling
to $b$-quarks it must have a dominant $H'$ (MSSM-like) component. Still, for
a large enough branching fraction into $A_1 A_1$, the $H'-S$ mixing
angle \eqref{2.12e} in the heavy scalar Higgs sector must not be too small
and, notably, the coupling $g_{S A_1 A_1}$ in \eqref{2.17e} must be large.
Both of these conditions are satisfied for $\kappa\sim 1.65$, which is
required if a single state should generate a visible signal.

Suitable values of
$\lambda$, $\tan\beta$ and $\mu$ for the desired masses and mixings
are given by a benchmark point BP1 in Table~1. (Since the mass of the
second heavy Higgs state is near 1~TeV and not near 750~GeV, these values
deviate somewhat from the ones obtained in the previous section.
Radiative corrections of ${\cal O}(\kappa^2/4\pi^2 \sim 0.07)$ can require
corresponding readjustments of these values.)
 Since $\tan\beta$ is
$\sim 10$, the NMSSM-specific uplift of the Standard Model
like Higgs mass at low $\tan\beta$ is not available. Then the Standard Model
like Higgs mass of $\sim 125$~GeV requires large radiative corrections as
in the MSSM.

As stated above and discussed in~\cite{Domingo:2016unq}, the squark masses
and $A_t$ can be chosen such that flavour violating couplings of $A_1$
are suppressed. In order to generate simultaneously large enough radiative
corrections to the Standard Model like Higgs mass, both parameters have to
be relatively large in the multi-TeV range. Possible numerical values are
also indicated in Table~1. The remaining NMSSM specific parameters $A_\lambda$
and $A_\kappa$ are chosen small, such that the $BR(H_{SM}\to A_1A_1)$
(depending somewhat on $A_\lambda$) is below $5\times 10^{-4}$, and $M_{A_1}$
sufficiently close to $2m_\mu$ such that the total width of $A_1$ is
large enough, i.e. that its decay length $l$ at 375~GeV is small enough:
For the BP1 in Table~1 with $M_{A_1}\sim 211.3$~MeV we get $\Gamma_{tot}(A_1)
\sim 1.74\times 10^{-13}$~GeV and $l \sim 2$~m, for which we estimate
that $1-e^{-d/l} \sim 63\%$ of all $A_1$ decays take place before the EM calorimeters.
($d$ denotes the average distance to the calorimeter cells of $\sim 2$~m.)
For the production cross sections of the Higgs state $H_2$ at 750~GeV we
find from SuShi\_1.5.0~\cite{Harlander:2012pb}
(at NNLO with MMHT2014 PDFs) $\sigma_{ggF}(H_2)\sim 4.8$~fb,
$\sigma_{bbH}(H_2)\sim 36.8$~fb, and from NMSSMTools we find
$BR(H_2\to A_1A_1)\sim 0.51$ with a total width of $H_2$ of $\sim 7$~GeV.
Together with a $BR(A_1\to \gamma\gamma) \sim 0.74$ we obtain a signal
rate of $\sim 4.6$~fb.  This signal rate remains to be multiplied by
an acceptance $Acc(\gamma)$ for the diphotons to simulate a single
photon in the detector. This issue will be discussed for all scenarios
in section~3.3; for the time being the signal rates appear with a
factor $Acc(\gamma)$ in Table~2.

If we assume a slightly smaller value of $M_{A_1}\sim 210.5$~MeV,
$\Gamma_{tot}(A_1)$ decreases to $\sim 1.65\times 10^{-13}$~GeV leading
to $l \sim 2.2$~m, reducing the percentage of decays before the EM
calorimeters to $\sim 60\%$ and hence the
signal rate by $\sim 10\%$.

Scenarios with smaller values of $\kappa$ are also possible. Then,
however, the reduced coupling $X_d$ of $A_1$ to leptons is smaller
(see~\eqref{2.19e}),
and the total $A_1$ width decreases. Hence the decay length increases,
and a smaller fraction of $A_1$'s decay before 2~m. This loss can be compensated
for if two states $H_2$ and $H_3$ with large production cross sections
and branching fractions into $A_1A_1$ contribute to the signal. 

The
benchmark point BP2 is of this type, where we take $\kappa=0.75$,
nearly (but not quite) small enough for the absence of a Landau singularity
below the GUT scale. For $M_{A_1}\sim 211.1$~MeV the total $A_1$ width
is $\sim 6.5\times 10^{-14}$~GeV, leading to $l \sim 5.5$~m. We estimate
that then only $\sim 30\%$ of all $A_1$ decays take place before the
EM calorimeter cells.
On the other hand, two Higgs states $H_2$ and $H_3$ with masses near
730~GeV and 762~GeV contribute to the signal. Both are strong mixtures
of the pure MSSM-like and singlet-like states. For $H_2$, the production
cross section is $\sigma_{ggF+bbH}(H_2)\sim 69.4$~fb, and
$BR(H_2\to A_1A_1)\sim 0.66$. For $H_3$, the production
cross section is $\sigma_{ggF+bbH}(H_3)\sim 54.3$~fb, and
$BR(H_2\to A_1A_1)\sim 0.53$.
Together with a $BR(A_1\to \gamma\gamma) \sim 0.73$ we obtain a signal
rate of $\sim 3.7$~fb times $Acc(\gamma)$, as shown in Table~2.

\begin{table}[t!]
\begin{center}
\begin{tabular}{|c|c|c|c|c|} \hline
Couplings and mass parameters &
BP1 &
BP2 &
BP3 &
BP4  \\\hline
$\lambda$ & 0.528 & 0.212 & 0.0332 & 0.0644 \\\hline
$\kappa$  & 1.65 & 0.75 & 0.121 & 0.215 \\\hline
$\tan\beta$ & 9.57 & 16.8 & 15.5 & 14.5 \\\hline
$\mu$ (GeV) & 138.5 & 101.1 & 102.3 & 111.3   \\\hline
$A_\lambda$ (GeV) & 32.2 & 15.6 & 0.0 & 0.0   \\\hline
$A_\kappa$ (GeV) & 1.16 & $7.67\times 10^{-2}$ & $-4.69\times 10^{-4}$ &
 $-1.49\times 10^{-3}$  \\\hline
$M_\text{squarks}$ (TeV) & 6  & 7.5 & 2 & 3 \\\hline
$A_t$ (TeV)& -3.48 & -3.95 & 3 & 3  \\\hline
\end{tabular}
\end{center}
\caption{Parameters for the four benchmark points. The soft Susy
breaking gaugino masses are $M_1=600$~GeV (500~GeV for BP1), $M_2=1$~TeV, $M_3=3$~TeV,
all squarks are assumed degenerate, and all slepton masses are 300~GeV
(with vanishing trilinear couplings). More digits (for all parameters) than shown here are
necessary in order to reproduce $M_{A_1}$ given in the Table~2 below,
and in order to obtain $C_A$ suffiently small for BP1 and BP2, see the text.}
\label{tab:1}
\end{table}

\begin{table}[t!]
\begin{center}
\begin{tabular}{|c|c|c|c|c|} \hline
 &
BP1 &
BP2 &
BP3 &
BP4  \\\hline
$M_{H_1}$ (GeV) & 122.1 & 124.3 & 123.7 & 122.2 \\\hline
$M_{H_2}$ (GeV)  & 750 & 730 & 744 & 740 \\\hline
$M_{H_3}$ (GeV) & 1003 & 762 & 750 & 750 \\\hline
$M_{A_1}$ (MeV) & 211.3 & 211.1 & 548.7 & 510.3   \\\hline
$M_{A_2}$ (GeV) & 763 & 747 & 748 & 745   \\\hline
$M_{H^\pm}$ (GeV) & 757 & 749 &752 & 749  \\\hline
$\sigma_{ggF}(H_2)$ (fb) & 4.8  & 2.2 & 1.7& 1.9 \\\hline
$\sigma_{bbH}(H_2)$ (fb) & 36.8 & 67.2 & 44.7 & 44.9 \\\hline
$\sigma_{ggF}(H_3)$ (fb) & 0.1  & 1.8 & 2.0 & 1.9 \\\hline
$\sigma_{bbH}(H_3)$ (fb) & 0.2  & 52.5 & 54.3 & 44.3 \\\hline
$BR(H_2\to A_1A_1)$ & 0.51  & 0.66 & 0.082& 0.21 \\\hline
$BR(H_3\to A_1A_1)$& 0.72 & 0.53 & 0.048 & 0.16  \\\hline
$BR(A_1\to \gamma\gamma)$& 0.74 & 0.73 & 0.72 & 0.66  \\\hline
$\Gamma_{tot}(A_1)$ ($10^{-13}$~GeV)& 1.74 & 0.65 & 7500 & 19000  \\\hline
$l(A_1)$ at 375~GeV & 2~m & 5.5~m & 0.18~mm & 0.08~mm  \\\hline
Signal cross section (fb) & 4.6 $\times Acc(\gamma)$ & 3.7 $\times Acc(\gamma)$
 & 3.4 $\times Acc(\gamma)$ & 6.7 $\times Acc(\gamma)$  \\\hline
\end{tabular}
\end{center}
\caption{Higgs masses, production cross sections and branching fractions
for the 4 benchmark points. For the points BP1 and BP2, the signal cross
section takes into account losses from $A_1$ decays beyond 2\,m according to a factor
$\left(1-e^{-2/l}\right)^2$ with $l$ in m. For the points BP3 and BP4 the
$BR(A_1\to \gamma\gamma)$ includes the $BR(A_1\to 3\pi^0)$. $Acc(\gamma)$
denotes the acceptance for the di- or multiphotons from two pseudoscalars
to simulate two single photons in the detector as discussed in section~3.3.
$M_{H_1}$ allows
for a theoretical error of $\sim 3$~GeV. For BP4, $\Gamma_{tot}(A_1)$
and $l(A_1)$ are the unreliable parton level results (at NLO).}
\label{tab:2}
\end{table}

\subsection{$M_{A_1}$ at 510-550~MeV}

The partial widths of a light pseudoscalar in this mass range can be
estimated employing two complementary approaches. To begin with one can
ask what one would obtain within the parton model, extrapolated into
the nonperturbative domain of QCD. First, 
for a reduced coupling of $A_1$ to leptons $X_d \sim 0.1$ as considered
below, the partial width of $A_1$ into muons can still be computed
reliably and is
\beq
\Gamma(A_1\to \mu^+\mu^-) \sim 5\times 10^{-11}\ \text{GeV}\; .
\eeq
The loop induced partial width of $A_1$ into $\gamma\gamma$ is $\sim 
4\times 10^{-15}$~GeV and hence negligibly small.
At NLO QCD the partial width of $A_1$ into strange quarks is
about $5\times 10^{-10}$~GeV and the loop induced width into gluons of the same
order as the width into $\mu^+\mu^-$. These widths can only be
rough estimates, however.

An alternative approach is to
consider the case $M_{A_1}\approx 550$~MeV, where one can expect
that $A_1$ mixes with the $\eta$ meson with a mass of 547.85~MeV.
(The possible r\^ole of $\eta$ for the decays of a light
pseudoscalar has been indicated earlier in~\cite{Dobrescu:2000jt}
without quantitative statements, however.) Mixing with the $\pi^0$ meson
of a lighter $A_1$ with $M_{A_1}\approx 135$~MeV has been
considered in~\cite{Domingo:2016unq}, where Partial Conservation
of Axial Currents (PCAC) or the sigma model for light mesons is
employed in order to determine the off-diagonal element of the
$A_1$-meson mass matrix; the same formalism will be used here for $A_1-\eta$
mixing for $M_{A_1}\approx 550$~MeV.

First we discuss this latter case, where the results can be
considered as more reliable. Only subsequently we turn to the case $M_{A_1}\approx
510$~MeV, motivated by the alleviation of constraints from radiative
$\Upsilon(1S)$ decays in this mass range, see below. There, however,
estimates of partial widths of $A_1$ are more speculative.

For $M_{A_1}\approx 550$~MeV, the relevant
mass matrix of the $A_1-\eta$ system reads in the basis
$(A_1,\eta)$
\beq
\frac{1}{2}\left(\ba{cc} M_{A_1}^2 & \delta m^2_{A_1 \eta}  \\ 
\delta m^2_{A_1 \eta}& m^2_\eta \ea\right)\; .
\eeq
For a small mixing angle $\theta$,
\beq
\theta \sim \frac{\delta m^2_{A_1 \eta}}{M_{A_1}^2-m^2_\eta}\ll 1\;,
\label{3.5e}
\eeq
the eigenstate $A_1'$ contains a small $\eta$ component:
$A_1' \sim A_1 + \theta \eta + ...$\; . For the partial widths of $A_1'$
one obtains then
\beq
\Gamma(A_1'\to X) \simeq \Gamma(A_1\to X) + \theta^2\Gamma(\eta\to X)\; .
\eeq
The dominant $\eta$ decays are~\cite{Agashe:2014kda}
\bea
\label{3.7e}
BR(\eta \to \gamma\gamma)&\sim& 39\%\;,\quad BR(\eta \to 3\pi^0)\sim 33\%\;,\quad
BR(\eta \to \pi^+\pi^-\pi^0)\sim 23\%\;,\nn\\
\Gamma_{tot}(\eta)&\sim& 1.3\times 10^{-6}\ \text{GeV}\; .
\eea
Next we require that the $\eta$-induced decays into $\gamma\gamma$ or $3\pi^0$
of the eigenstate $A_1'$ dominate its width into $\mu^+\mu^-$, since we
ignore the unreliable widths of $A_1$ into strange quarks or gluons in
this subsection. (Since the latter decays can also generate $\gamma\gamma$
or $3\pi^0$ final states, this assumption is conservative.) This leads to
\beq
\theta^2 > \frac{\Gamma(A_1\to \mu^+\mu^-)}{\Gamma_{tot}(\eta)}\;,
\ \ \theta \gsim 6\times 10^{-3}\; .
\label{3.8e}
\eeq

In order to estimate the mixing matrix element $\delta m^2_{A_1 \eta}$
above we use, following~\cite{Domingo:2016unq}, PCAC. There one
introduces the SU(3)$_\text{axial}$ flavour currents $J^\mu_{A\; i}$
where $i$ denote the SU(3) generators. Assuming that $\eta$ is a pure
octet, $J^\mu_{A\; 8}$ satisfies
\beq
\partial_\mu J^\mu_{A\; 8} = f_\pi m_\eta^2 \eta
\eeq
with $f_\pi \sim 93$~MeV. At the quark level one has
\beq
\partial_\mu\left( \bar{s}\gamma^\mu\gamma_5 s\right) =
-\sqrt{\frac{2}{3}}\partial_\mu J^\mu_{A\; 8}
+\frac{1}{\sqrt{3}}\partial_\mu J^\mu_{A\; 0}
\eeq
where $J^\mu_{A\; 0}$ is the (anomalous) U(1)$_A$ current whose divergence
involves the $\eta'$ meson. Using these relations, one can re-write the
coupling of $A_1$ to strange quarks in the Lagrangian (proportional to
the corresponding Yukawa coupling $X_d m_s/v$)
\beq
\frac{-i m_s X_d}{\sqrt{2} v} A_1 \bar{s}\gamma_5 s
=-\frac{X_d}{2\sqrt{2} v} A_1 \partial_\mu \left(\bar{s}\gamma^\mu\gamma_5 s\right)
=\frac{X_d}{2\sqrt{3} v} A_1 \partial_\mu  J^\mu_{A\; 8} +...
=\frac{X_d f_\pi m_\eta^2}{2\sqrt{3} v}\eta A_1 +...
\label{3.11e}
\eeq
where we have dropped the terms $\sim \partial_\mu J^\mu_{A\; 0} \sim \eta'$. From
\eqref{3.11e} one can read off
\beq
\delta m^2_{A_1 \eta} = \frac{X_d f_\pi m_\eta^2}{2\sqrt{3} v}\; .
\eeq
Then the request \eqref{3.8e} becomes, again for $X_d \sim 0.1$ and using \eqref{3.5e},
\beq
\left|M_{A_1}-m_\eta\right| < 10^{-3} m_\eta \sim 0.5~\text{MeV}\; .
\label{3.13e}
\eeq

This estimate can be refined by including mixing with the $\eta'$ meson,
the anomalous U(1)$_A$ current $J^\mu_{A\; 0}$ and the loop-induced
coupling of $A_1$ to $F_{\mu\nu}\widetilde{F}^{\mu\nu}$, where $F_{\mu\nu}$
is the QCD field strength\footnote{We thank F. Domingo for providing
us with his studies of this issue.}. The additional contribution to
$\delta m^2_{A_1 \eta}$ leads to a replacement of the right hand side of
\eqref{3.13e} by $\sim 1$~MeV.

Assuming such a small $A_1 - \eta$ mass difference, the decay length of
$A_1$ is below a mm, and its branching fractions are the ones of $\eta$
given in \eqref{3.7e} above. For $M_{A_1}>M_K$, constraints from rare $K$~decays are no
longer relevant. However, since $A_1$ has couplings to $b$-quarks
$\sim X_d$, constraints from the search for the radiative
decays $\Upsilon(1S) \to \gamma\, \eta$ by CLEO in~\cite{Athar:2007hz} apply
(and are more
relevant than searches for $\Upsilon(1S) \to \gamma A_1 \to \gamma \mu^+\mu^-$).

The $BR(\Upsilon(1S)\to\gamma A_1)$ can be obtained from the Wilczek
formula~\cite{Wilczek:1977pj,Hodgkinson:2008ei}
\bea
\label{3.14e}
\frac{BR(\Upsilon(1S)\to\gamma A_1)}{BR(\Upsilon(1S)\to\mu^+\mu^-)} &=&
\frac{G_F m_b^2 X_d^2}{\sqrt{2}\pi\alpha_{em}}
\left(1-\frac{M_{A_1}^2}{M_{\Upsilon(1S)}^2}\right)\times F\, ,\ \text{hence}
\nonumber\\
BR(\Upsilon(1S)\to\gamma A_1) &\sim & 1.03\times 10^{-4}\times X_d^2
\eea
where $BR(\Upsilon(1S)\to\mu^+\mu^-)\sim 2.48\%$ and $F$ is a correction factor
$\sim 0.5$.
The upper bound of CLEO~\cite{Athar:2007hz} on $BR(\Upsilon(1S)\to\gamma \eta)$ is
$1.0\times 10^{-6}$ at the 90\% CL level, or $1.3\times 10^{-6}$ at the
95\% CL level. Applying this bound to the $BR(\Upsilon(1S)\to\gamma A_1)$,
\eqref{3.14e} gives
\beq
X_d \lsim 0.11
\eeq
as used above. From \eqref{2.19e} one finds that $\kappa$ must then also
be quite small, leading to relatively small branching fractions of the
heavy Higgs states $H_2$ and $H_3$ into $A_1A_1$. Hence both of these states
should contribute to the signal.

Next we consider the $A_1$ decays induced by its mixing with $\eta$ where
$\eta$ decays as in \eqref{3.7e}.  The decays into
$\pi^0 \pi^+\pi^-$ give diphotons plus muons, but due to the escaping
neutrinos this final state will not allow to reconstruct the masses of
the original resonances near 750~GeV. In addition to
the $A_1\to \gamma\gamma$ mode, the $A_1\to 3\pi^0$ mode leads to
photon jets. The compatibility of such photon jets with a single photon
signature in the detectors has been discussed in detail in~\cite{Dasgupta:2016wxw}.
In particular, due to the enhanced probability for photon conversions into $e^++e^-$
in the inner parts of the detectors, such scenarios can be distinguished from
single photons (or even diphotons) once more events are available. Adding both
modes, about $72\%$ of all $A_1$ decays lead to di- or multi-photons. The
resulting signal cross section remains to be multiplied by the acceptamce
$Acc(\gamma)$ for the di- or multi-photons to fake a single photon
discussed in section~3.3.

The parameters, masses, branching fractions, production and signal cross
sections of a corresponding benchmark point BP3 are shown in Tables~1 and~2.

If $M_{A_1}$ differs by a few tens of MeV from the $\eta$ mass it becomes more
difficult to estimate its decays; its mixing angle with the on-shell $\eta$
meson using PCAC as above becomes tiny. Its Yukawa couplings to Standard Model
fermions are obtained through mixing with the (heavy) MSSM-like pseudoscalar
$A_{MSSM}$. At the parton level and for $\tan\beta \sim 10-15$, the
relative couplings squared and hence the corresponding widths of $A_1$ are
dominantly into $s\bar{s}$ ($\approx 87\%$),
into $gg$ via top quark loops ($\approx 5\%$), and into $\mu^+\mu^-$ ($\approx 8\%$).
The hadronic or $\gamma\gamma$ decays of $A_1$ can then be considered as being
mediated by the CP-odd isospin and color singlet interpolating composite
fields $s\bar{s}$ and $F_{\mu\nu}\widetilde{F}^{\mu\nu}$.
Both are known components of the $\eta$ wave function in Fock space,
and the most reasonable assumption is that their hadronisation
(decays into physical hadrons and $\gamma\gamma$) proceeds again with branching
fractions similar to the ones of $\eta$. 

The partial width for the sum of these decays of $A_1$ is less clear,
however. It is relevant, since it competes with the
width of $A_1$ into $\mu^+\mu^-$ and determines consequently the
branching fraction $BR(A_1\to \text{hadrons\ or}\ \gamma\gamma)$ via the above
interpolating fields relative to the $BR(A_1\to \mu^+\mu^-)$.
Since the widths for the above mentioned decays of
$\eta$ into $\gamma\gamma$ or pions are small (being electromagnetic or
suppressed by isospin), one must assume that the widths for the decays
of $A_1$ via the above interpolating fields are also smaller than estimated
from the couplings squared at the parton level as at the beginning of
this section. A quantitative statement
is difficult, however, without a nonperturbative evaluation of the
relevant matrix elements between physical states. 

On the other hand, the sum of the couplings squared of $A_1$ to $s\bar{s}$ or $gg$
and hence the sum of the partial widths of $A_1$ into $s\bar{s}$ or $gg$
(for both of which $\eta$-like branching fractions are assumed) is considerably
larger than into $\mu^+\mu^-$: At NLO one has $BR(A_1\to s\bar{s}\ \text{or}\ gg)
\sim 0.92$. Hence, reducing the sum of the partial widths of $A_1$ into
$s\bar{s}$ or $gg$ by a factor 1/10 leaves us still with a dominant
$BR(A_1\to s\bar{s}\ \text{or}\ gg)\sim 0.55$.

In the scenario where
$M_{A_1}$ differs by a few tens of MeV from the $\eta$ mass we will make the
assumption that the reduction of the width of the decays
$A_1\to \gamma\gamma\ \text{or\ hadrons}$ is not too dramatic, i.e. the
relevant branching fractions of $A_1$ can be parametrized as
\beq
BR(A_1\to \gamma\gamma,3\pi^0,\pi^0 \pi^+\pi^-) \sim F_A \times
BR(\eta\to \gamma\gamma,3\pi^0,\pi^0 \pi^+\pi^-)
\label{3.16e}
\eeq
where the factor $F_A$ is not too small $(\gsim 0.1)$.

Let us have another look at the searches by CLEO which
were performed separately for the $\eta\to 3\pi^0$, $\eta\to \pi^0
\pi^+ \pi^-$ and $\eta\to \gamma\gamma$ final states. The windows for the
invariant masses were chosen differently for different final states,
$M_{3\pi^0}\gsim 475$~MeV and $M_{\pi^0\pi^+ \pi^-} \gsim 515$~MeV.
($M_{\gamma\gamma}$ is fitted to a double Gaussian function centered
at $M_\eta$.) No candidates were found in the (background free)
$3\pi^0$ and $\pi^0\pi^+ \pi^-$ final states, but two events in
$\pi^0\pi^+ \pi^-$ with $M_{\pi^0\pi^+ \pi^-}\sim 510$~MeV just below
the $M_{\pi^0\pi^+ \pi^-}$ window. Also a mild excess of events for
$M_{\gamma\gamma}\sim 510$~MeV is observed. These events are not
numerous enough to allow for the claim of a signal, but we conclude that the
$\pi^0\pi^+ \pi^-$ and $\gamma\gamma$ final states do not lead to
stronger upper limits on $BR(\Upsilon(1S)\to\gamma A_1)$
for $M_{A_1}\sim 510$~MeV than the limit on $BR(\Upsilon(1S)\to\gamma \eta)$
from the remaining $3\pi^0$ final state. After
translating the 90\%~CL upper limit from the latter final state into
a 95\%~CL upper limit, we find from only the $3\pi^0$ final state
\beq
BR(\Upsilon(1S)\to\gamma \eta)\times 0.33 \lsim 3.8\times 10^{-6}
\label{3.17e}
\eeq
where 0.33 is the $BR(\eta \to 3\pi^0)$. For $A_1$, assuming $\eta$-like decays,
this upper bound becomes
\beq
BR(\Upsilon(1S)\to\gamma A_1)\times 0.33\times F_A \lsim 3.8\times 10^{-6}
\label{3.18e}
\eeq
Combining \eqref{3.18e} and \eqref{3.14e} gives
\beq
X_d \lsim 0.19/\sqrt{F_A}
\label{3.19e}
\eeq
and, from \eqref{2.19e},
\beq
\kappa \lsim 0.21/\sqrt{F_A}\; .
\label{3.20e}
\eeq
Since smaller $F_A$ alleviates the constraint on $\kappa$, we should study
its impact on the signal rate. First, the branching fractions of the heavy
Higgs states induced by the coupling $g_{S A_1 A_1}$ in \eqref{2.17e} behave
roughly like $g_{S A_1 A_1}^2 \sim \kappa^4 \lsim 0.002\; F_A^{-2}$, i.e. smaller $F_A$
allows for larger branching fractions. On the other
hand, by assumption (see \eqref{3.16e}) the branching fractions of $A_1$ into $\gamma\gamma$
or $3\pi^0$ are proportional to $F_A$. Hence the factors of $F_A$ cancel approximatively
in the final signal rate. (Depending on the other parameters we found,
however, that smaller $F_A$ can lead to a decrease of the signal rate
if the heavy Higgs branching fractions increase somewhat less than indicated above.)

For a rough estimate we have constructed a benchmark point BP4 with
$M_{A_1}\sim 510$~MeV. It has $X_d \simeq 0.206$ and satisfies
the CLEO constraints for $F_A < 0.87$. Its parameters are given in Table~1.
For the branching fractions of $A_1$ into $\gamma\gamma$ or $3\pi^0$ we
assume $F_A \sim 0.87$ which gives, using the corresponding branching
fractions of $\eta$, $BR(A_1\to \gamma\gamma$ or $3\pi^0) \sim 0.87\times(0.39+0.33)
\simeq 0.63$. Together with the production cross sections and branching
fractions of the Higgs states $H_2$ and $H_3$ in Table~2 we obtain
finally a signal cross section of $\sim 6.7$~fb times $Acc(\gamma)$.

\subsection{The spread in rapidity of multiphotons}

The probability for a di- or multiphoton system to fake a single photon
depends to a large extent on its angular spread. In the context of a decay
of the Standard Model Higgs boson into light pseudoscalars (which decay into
diphotons) this has been discussed in some detail in \cite{Draper:2012xt}.
One has to consider the fineness in rapidity $\eta$ of the strips of the first layer
of the electromagnetic calorimeter (EM), ranging from 0.003 to 0.006 (depending
on $\eta$) for ATLAS \cite{Aad:2010sp}. A relevant quantity for an event is the fraction
of the total deposited energy in a single strip, and which
fraction is deposited in the adjacent strips \cite{Draper:2012xt}. In the
case of diphotons, the relevant criterium is then the distribution of
$\Delta \eta$ between the two photons. In \cite{Draper:2012xt} it is argued (and used in
\cite{Aparicio:2016iwr}) that only for $\Delta \eta \lsim 0.0015$
prompt diphotons fake a single photon. (Including converted photons
does not seem to modify this estimate \cite{Draper:2012xt}.) This number
remains to be confirmed by the experimental collaborations, however,
and will depend on $\eta$ and the transverse energy of the individual
events in practice.

In the case of displaced vertices, as for our benchmark points BP1 and
BP2, the situation is more involved as discussed in \cite{Aparicio:2016iwr}:
diphotons from pseudoscalars decaying between the original vertex and
the EM are more collimated. Note that the signal cross sections given in
Table~2 (without the factor $Acc(\gamma)$) take already into account the
loss from pseudoscalars decaying inside or beyond the EM. The interplay
between these losses and $\Delta \eta$ for diphotons reaching the EM has
been studied in \cite{Aparicio:2016iwr}, but is beyond the scope of the
present paper. We content ourselves with the fact that $Acc(\gamma)$
for our benchmark points BP1 and BP2 is then definitively larger than
$Acc(\gamma)$ for pseudoscalars of a corresponding mass of $\sim 200$~MeV
which decay promptly. To this end we studied the distribution of
$\Delta \eta$ for diphotons from promptly decaying pseudoscalars with the
help of a simulation based on MadGraph/\-MadEvent~\cite{Alwall:2011uj} 
including Pythia~6.4 \cite{Sjostrand:2006za}, where the pseudoscalars
originate from a 750~GeV Higgs state. The resulting distribution is
shown in red in Fig.~1.

\begin{figure}[ht!]
\begin{center}
\vspace*{-6cm}
\includegraphics[scale=0.9,trim= 0 50 0 330, clip]{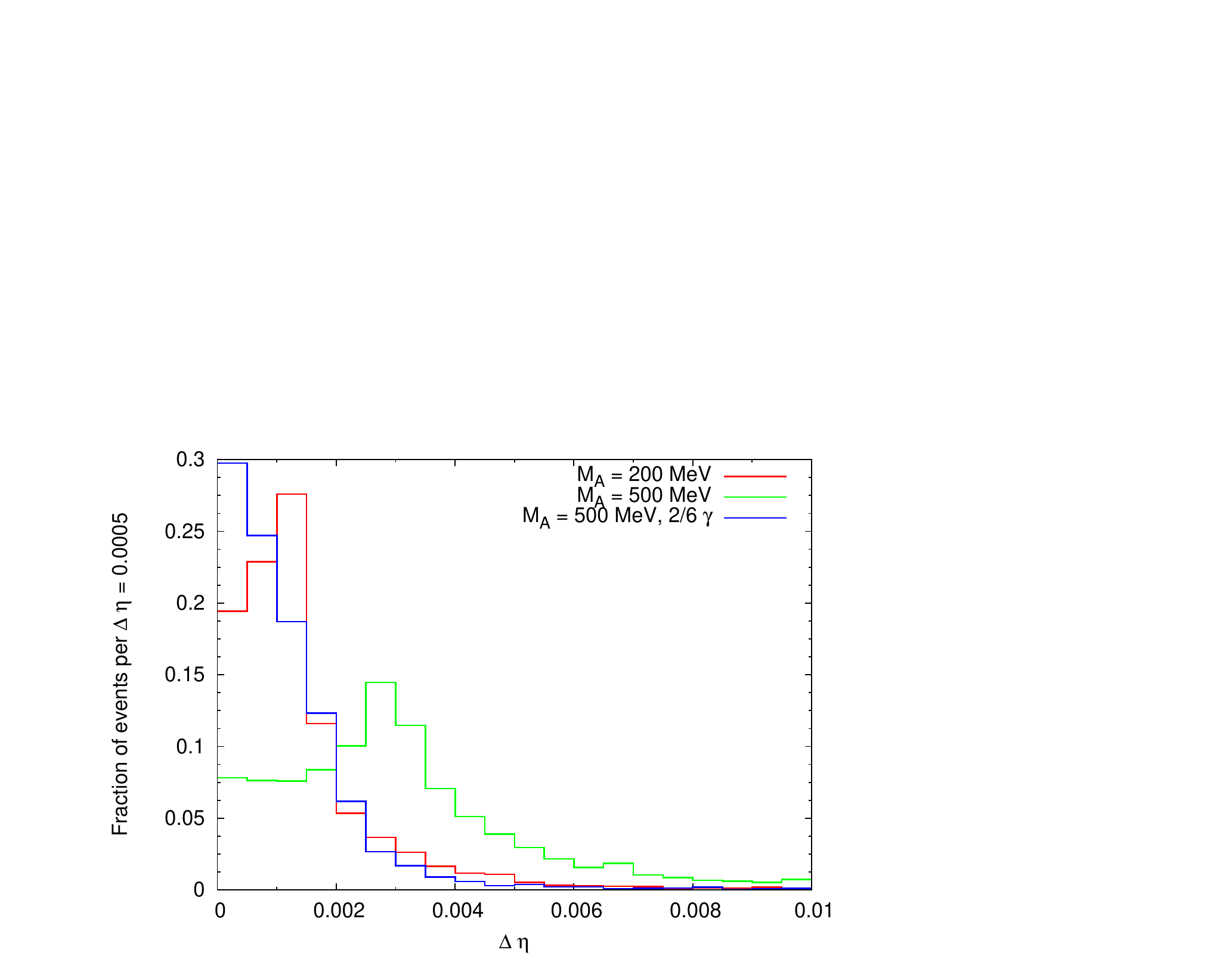}
\end{center}
\caption{Distributions of $\Delta \eta$ of diphotons from promptly
decaying pseudoscalars with a mass of 200~MeV (red) and 500~MeV
(green). 500~MeV, 2/6 $\gamma$ (blue) denotes $\Delta \eta$ between the two leading among
all 6 photons from the decays $A_1 \to 3\pi^0\to 6\gamma$.}
\label{fig:1}
\end{figure}

We find that $70\%$ of these diphotons satisfy $\Delta \eta \lsim 0.0015$
(in rough agreement with \cite{Aparicio:2016iwr}),
and correspondingly more due to the displaced vertices for our benchmark
points BP1 and BP2. A rough estimate for $Acc(\gamma)$ for the diphotons
from both pseudoscalars to fake a single photon would then be 
$\approx 75\%$. We underline, however, that the actual number of signal events
responsible for the observed excesses is small, and statistical fluctuations
of quantities like $\Delta \eta$ can be correspondingly large.

Turning to the scenarios BP3 and BP4, displaced vertices are no longer relevant.
However, here we expect about equal fractions of both diphoton and 6-photon final states, the
latter from $A_1\to 3\pi^0$. For $M_{A_1}\sim 500$~MeV, the spread in
$\Delta \eta$ for the diphotons is obviously larger, as shown in Fig.~1 in
green. Here only $\sim 25\%$ of all diphotons satisfy $\Delta \eta \lsim 0.0015$.

For the 6-photon final state we note that the angular
spread between the photons should be smaller, as the total invariant
mass of the system must remain the same and less energy is available
for momenta transverse to the principal axis. Since the relevant quantity
in the EM calorimeters is the spread of the deposited energy, we concentrate
here on $\Delta \eta$ between the two most energetic photons resulting from
a single $A_1\to 3\pi^0$ decay. It is shown
in blue in Fig.~1, and we find that in $\sim 75\%$ of all cases these are
closer than $\Delta \eta \sim 0.0015$. This number is suggestive, but it is not clear whether
it coincides with the fraction of 6-photon final states faking a single photon;
due to the more homogenous distribution of the deposited energy this
fraction could even be larger.
If we use it as it is, the $Acc(\gamma)$ for the di- or multiphotons
from both pseudoscalars to fake simultaneously a single photon becomes
$Acc(\gamma) \approx 25\%$.
Moreover many of 6 photons will convert, and the acceptance of
such events remains to be studied by the experimental collaborations.
Again, statistical fluctuations can be large as long as the number of signal
events is as low as at present.

\section{Summary and Conclusions}

We have shown that the excess of events in the diphoton final state
near 750~GeV observed by ATLAS and CMS can be explained within
a fairly standard supersymmetric extension of the Standard Model,
the NMSSM, without invoking new particles like additional vector-like
quarks and/or leptons. 
The signal cross sections are not very large, but may be sufficient
to explain the observed excesses.

The corresponding processes differ, however,
from what has been proposed in most of the literature up to now:
Except for the scenario BP1 (with $\kappa\sim 1.65$), two resonances
nearby in mass which
share the properties of the additional CP-even scalars of the NMSSM
 are responsable for the signal cross section.
Their components proportional to the MSSM-like scalar $H'$ lead to
enhanced couplings to $b$-quarks implying sizeable production
cross sections via associated production with $b$-quarks, whereas their
components proportional to the singlet-like scalar $S$ lead to sizeable
branching fractions into two pseudoscalars $A_1 A_1$.
These scenarios are not in tension
with the upper limit from CMS on the diphoton cross section obtained at
the run~I at 8~TeV~\cite{Khachatryan:2015qba}; note that the $bbH$ cross
sections increase somewhat faster with the c.m. energy than $ggF$.

We note that constraints from other decay modes of the scalars with
masses of about 750~GeV are satisfied: Upper bounds on signal cross
sections into other final states -- quark pairs, lepton pairs and
electroweak gauge bosons -- are discussed 
in~\cite{Knapen:2015dap,Low:2015qep,Gupta:2015zzs}.
These bounds are obeyed given the relatively
large branching fractions into $A_1 A_1$ and finally into
di- or multi-photon final states of the heavy scalars in our scenario, which
do not require excessive production cross sections.

Four different scenarios have been discussed, which differ in the
properties and masses of the light pseudoscalars $A_1$ and the heavy
Higgs states:

1) For the benchmark points BP1 and BP2, the mass $M_{A_1}\sim 211$~MeV
is just below twice the muon mass. Then the branching fraction of $A_1$
into diphotons
is large enough for a satisfactory signal rate. In the case of the BP1 with
$\kappa\sim 1.65$, a single heavy Higgs state (still a mixture of the
MSSM-like and singlet-like states) with a width of $\sim 7$~GeV is
sufficient for a signal. In the case of the BP2 with a more modest value for
$\kappa\sim 0.75$, two nearby heavy Higgs states, both with a width of
$\sim 6$~GeV, are responsible for the signal.
For both BP1 and BP2, the dominant constraints from low energy
experiments originate from $K$~decays involving loop-induced flavour
changing vertices of $A_1$; it must be assumed that these are cancelled
by suitable choices of the SUSY breaking parameters.

2) For the benchmark points BP3 and BP4 it is assumed that $A_1$ shares
its branching fractions with the $\eta$ meson. In the case of BP3 with
$M_{A_1}\sim 549$~MeV this is guaranteed by $A_1-\eta$ mixing, estimated
with the help of the PCAC formalism. In the case of BP4 with
$M_{A_1}\sim 510$~MeV, estimates of the $A_1$ partial widths are on less
solid ground. We assumed that the non-leptonic decays of $A_1$ proceed
via $s\bar{s}$ or $gg \sim F_{\mu\nu}\widetilde{F}^{\mu\nu}$ interpolating
fields which, in turn, hadronise (decay) again similar to the $\eta$
meson. We showed, however, that sizeable reductions of the corresponding
partial widths with respect to the decays of $A_1$ into $s\bar{s}$ or $gg$
by, e.g., a factor 1/10, would not invalidate this
scenario. For both BP3 and BP4 two nearby heavy Higgs states with widths
of $\sim 1.5-2$~GeV are responsible for the signal. For this range of
$M_{A_1}$ the dominant constraints from low energy
experiments originate from searches for radiative $\Upsilon$ decays into
$\gamma + \eta$ by CLEO. These lead to upper bounds on the coupling of
$A_1$ to down-type quarks and leptons and, as we have shown, on $\kappa$.

Interestingly, the four scenarios have different features which allow
to distinguish them experimentally also from more ``conventional''
models:

1) For BP2, BP3 and BP4 the signal originates from two resonances $H_2$
and $H_3$ close in mass, which can imitate a single wide resonance.
Of course, small variations of the parameters allow to vary the masses
of $H_2$ and $H_3$, the total signal rate, and to reshuffle the individual
signal rates of $H_2$ and $H_3$. With more events (and depending on the
actual mass difference) the two states could possibly be resolved.
A particular feature of BP1 is that the single resonance near 750~GeV
responsible for the signal has another large branching fraction of
$\sim 25\%$ into $Z+A_1$. With $A_1$ imitating a photon, one obtains
signals of the kind $Z+\gamma$ similar to the ones expected if a
750~GeV resonance decays into $\gamma\gamma$ via fermionic loops.

2) The $A_1$ decays differ considerably for the benchmark points. For BP1 and
BP2 the decay lengths of $A_1$ are macroscopic leading to measurable
displaced vertices if $A_1$ decays inside the calorimeters. BP2 corresponds
to an extreme case with a decay length of $\sim 5.5$~m, but a very large
signal cross section (before reducing it by the number of decays before
the EM calorimeters). For both BP1 and BP2, $A_1$ decays into diphotons.
However, due to the displaced vertices it will not be straightforward to
distinguish them from single photons via the
number of converted photons \cite{Dasgupta:2016wxw}.
Moreover, $A_1$ has branching fractions of $\sim 25-30\%$ into $e^++e^-$
leading to similar signatures as converted photons.
Also the opening angle between the two photons gets reduced for displaced
vertices increasing their acceptance as a single photon, see section~3.3.
For BP3 and BP4 the $A_1$ decay lengths are short,
but $A_1$ decays into diphotons or photon jets from $3\pi^0$. The latter
should lead to a very large proportion to ``converted photons''; additional
potentially relevant observables like EM shower shapes have also been discussed
in~\cite{Dasgupta:2016wxw}. Our results for $M_{A_1} \sim 500$~MeV for
$\Delta \eta$ between diphotons or the two leading photons from jets from $3\pi^0$
may be useful here.
Finally, for BP3 and BP4 $A_1$ has branching
fractions into muon pairs (of $6-8\%$ at the parton level)
which could be used for alternative signals, once more events are obtained.

Hence, if the excess of events continues, several observables can be used
to verify/test/exclude the scenarios discussed here.

Finally we recall that the origin of the light pseudoscalar $A_1$ in the
NMSSM is an approximate $R$-symmetry of the scalar potential, see the
small values of $A_\kappa$ and $A_\lambda$ of the benchmark points. For
BP1, BP2 and BP3, $M_{A_1}$ has to coincide accidentially with specific
values $2\,m_\mu$ or $m_\eta$. For the BP4 $M_{A_1}$ is actually
less constrained (unless one intends to fit the events near
$M_{A_1}\sim 510$~MeV observed by CLEO as we did), but in this scenario
the $A_1$ decays are less understood theoretically. Although
the approximate $R$-symmetry
at the weak (or SUSY) scale is not preserved by radiative corrections,
we content ourselves in the present paper with the mere fact that such
a scenario would allow to explain the events. Work on an
$R$-symmetric extension of the NMSSM explaining a light pseudo-Goldstone
boson naturally is in progress.

\section*{Acknowledgements}

The authors acknowledge
the support of France Grilles for providing cloud computing resources on the
French National Grid Infrastructure and
thank J.~Jaeckel and  M.~Spira for helpful discussions,
and notably F.~Domingo for communication on possible
$A_1-\eta$ mixings. U.E. acknowledges support from the European Union Initial
Training Networks Higgs\-Tools (PITN-GA-2012-316704), INVISIBLES (PITN-GA-2011-289442),
the ERC advanced grant Higgs@LHC, and from
the grant H2020-MSCA-RISE-2014 No. 645722 (NonMinimalHiggs).


\end{document}